%% file: fpga_asic_ising_codesign.tex
\documentclass[sigconf,9pt]{acmart}

\copyrightyear{2026}
\acmYear{2026}
\setcopyright{cc}
\setcctype{by}
\acmConference[ICCAD '26]{IEEE/ACM International Conference on Computer-Aided Design}{November 08--12, 2026}{San Jose, CA, USA}
\acmBooktitle{IEEE/ACM International Conference on Computer-Aided Design (ICCAD '26), November 08--12, 2026, San Jose, CA, USA}
\acmDOI{10.1145/3831252.3834138}
\acmISBN{979-8-4007-2873-0/2026/11}

\graphicspath{{./}{figure/}}

\usepackage{amsmath}
\usepackage{placeins}
\usepackage{multirow}
\usepackage{tabularx}
\usepackage{makecell}
\usepackage[font=small,skip=3pt]{caption}

\usepackage{enumitem}
\setlist[itemize]{leftmargin=*,labelsep=0.4em,topsep=2pt,itemsep=3pt,parsep=0pt}
\setlist[enumerate]{leftmargin=*,labelsep=0.4em,topsep=2pt,itemsep=3pt,parsep=0pt}

\begin{document}

\title{An FPGA-ASIC Co-Design Framework for Capacity-Constrained Physics-based Ising Chips}

\author{Ruihong Yin, Yue Zheng, Chaohui Li, Ahmet Efe, Abhimanyu Kumar,
  Ziqing Zeng, Ulya R. Karpuzcu, Sachin S. Sapatnekar, Chris H. Kim}
\affiliation{%
  \institution{Department of Electrical and Computer Engineering,
    University of Minnesota}
  \city{Minneapolis}
  \state{MN}
  \country{USA}
}
\email{{yin00473, zhen0631, li003135, efe00002, kumar663, zeng0083, ukarpuzc, sachin, chriskim}@umn.edu}

\renewcommand{\shortauthors}{Yin et al.}



\input{sections/0.abstract}

\maketitle

\input{sections/1.introduction}

\input{sections/2.background}
\input{sections/3.formulation}

\input{sections/4.architecture}
\input{sections/5.evaluation}

\input{sections/6.conclusion}

\begin{acks}
This work was supported in part by the DARPA Quantum-Inspired Classical
Computing (QuICC) program under AFRL contract FA8750-22-C-1034, and NSF grants
2230963 and 2142248.
\end{acks}

\clearpage
\bibliographystyle{ACM-Reference-Format-citorder}
\bibliography{refs}

\end{document}

%% file: sections/0.abstract.tex
\begin{abstract}
When a problem exceeds an analog Ising machine's spin capacity it cannot be
solved in one shot: a digital orchestration layer must iteratively decompose
the graph, clamp boundary spins, and deliver hardware-sized subproblems to the
solver.  As per-subproblem solve times reach the \textmu s-scale regime, this
digital layer (not the analog core) becomes the primary scalability
bottleneck.  On a 28~nm CMOS coupled-oscillator Ising chip
($T_{\mathrm{core}} \approx 77.5~\mu$s, 24~mW), a CPU orchestrator leaves the
solver idle for over 98\% of every iteration at $N{=}750$, and the gap
survives OpenMP and AVX2 optimization because it is dominated by data-dependent memory access rather than arithmetic throughput.
We argue that the orchestration layer is a first-class design object and
propose a sizing methodology for hybrid analog-digital Ising systems: given a
solver's core time, clock frequency, and target problem class, it gives an
analytical baseline for hardware parallelism and memory-bandwidth
dimensioning.  Guided by the sizing laws, an FPGA orchestration layer
co-located with the chip satisfies the pipeline condition with substantial
margin.  Across graph coloring, MaxCut, and SAT benchmarks it delivers
9.8$\times$--46.5$\times$ raw end-to-end time-to-solution (TTS) improvements
(14.8$\times$--65.9$\times$ on per-repeat runtime) over the optimized CPU
orchestrator, with device-level power reductions of 85$\times$--116$\times$.
\end{abstract}

%% file: sections/1.introduction.tex
\section{Introduction}\label{sec:introduction}

Many important combinatorial optimization problems, including Boolean
satisfiability (SAT), graph coloring, and MaxCut, can be expressed as Ising
Hamiltonians, making
them natural targets for specialized Ising
hardware~\cite{Lucas_2014,Prasad2005,JunDongCho1998}.  Physics-based analog
solvers suit these tasks: they search the energy landscape in continuous time
and solve hard discrete problems at low power.

\begin{figure}[t]
\centering
\includegraphics[width=\linewidth]{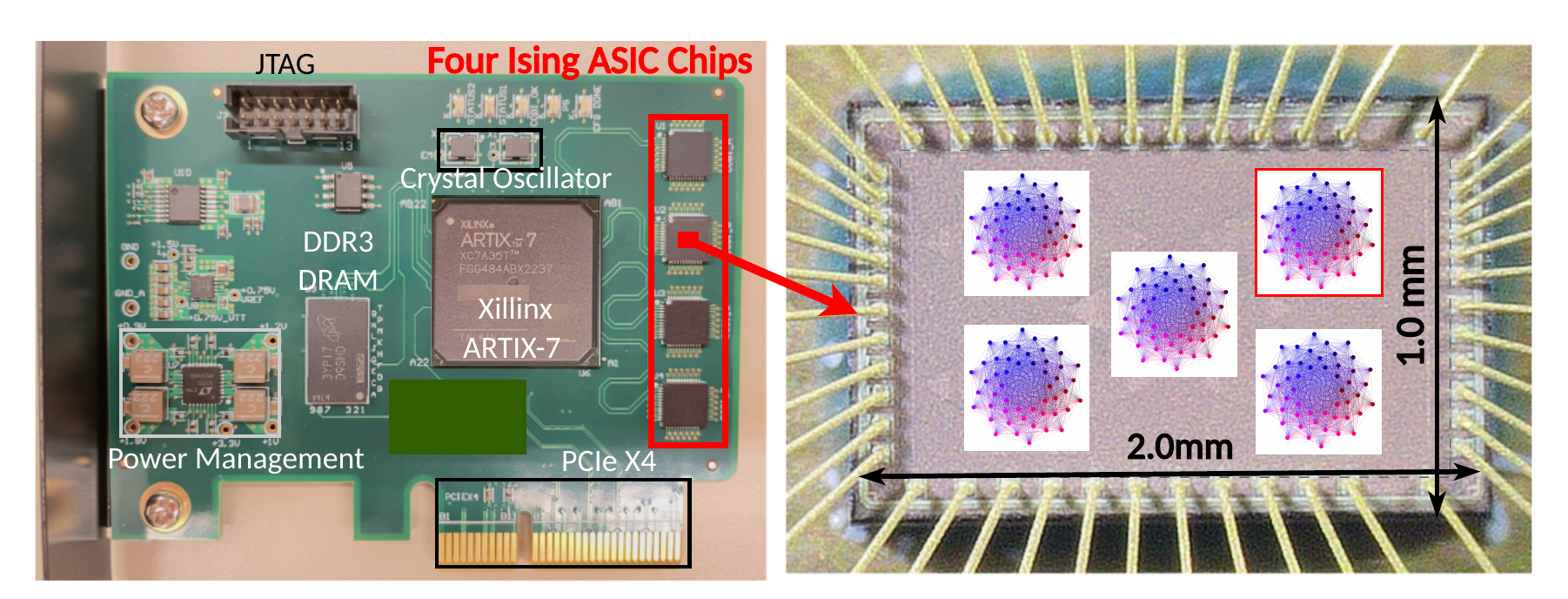}
\caption{(a) Custom PCIe card with FPGA host and four Ising ASIC chips, each
with five all-to-all connected oscillator cores. (b) Die photograph in 28~nm
CMOS.  Each core completes one 45-spin solve in $\approx$77.5~$\mu$s at 24~mW,
driven by the on-board FPGA via a 16-bit shared input bus and individual 1-bit
output lines.}
\label{fig:cobi_layout}
\end{figure}

Despite this promise, the main limitation of analog Ising hardware is often
not the solver itself but the size of the problem it can handle at once.  Here
\emph{capacity} means the maximum number of spins one core can process in a
single solve; each hardware spin implements one binary variable, so capacity
directly limits the problem size addressable in one shot.  Our target platform
(Fig.~\ref{fig:cobi_layout}) is a 28~nm CMOS coupled-oscillator Ising chip
whose five cores each solve at most a 45-spin all-to-all-coupled
subproblem.  A problem with more than 45 variables cannot fit on one core; it
must be decomposed into hardware-sized subproblems and solved iteratively.

\begin{figure}[t]
\centering
\includegraphics[width=\linewidth]{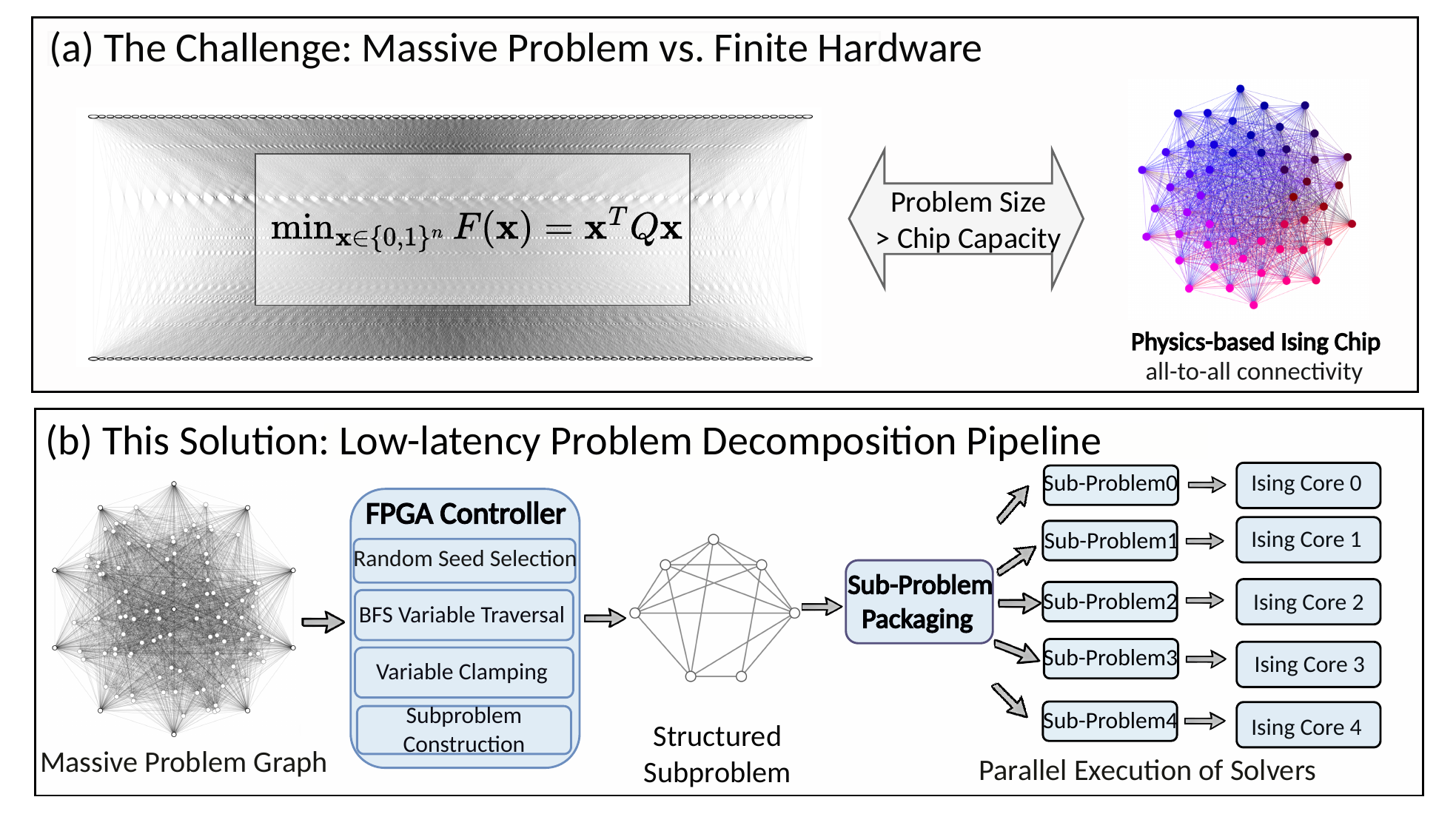}
\caption{Capacity mismatch and low-latency feeding pipeline.  When $N$ exceeds
core capacity, the FPGA decomposes the graph, clamps boundary variables,
constructs hardware-sized subproblems, and feeds the cores within the
$\mu$s-scale solve window.}
\label{fig:capacity_fpga_control}
\end{figure}

This creates a new system bottleneck: the digital orchestration layer, the
external controller that prepares and feeds subproblems to the chip.  Every
iteration it must select a 45-spin subset, fold in the influence of the
outside spins by clamping, build the clamped subproblem, and deliver it before
the current solve finishes.  One solve takes only $\approx$77.5~$\mu$s, and
missing that deadline is costly: a CPU orchestrator needs 2.15~ms per
iteration on a 300-spin instance ($27.7\times$ the solve window),
and the analog cores idle away more than 96\% of every iteration, below a 2\%
duty cycle at $N{=}750$.  The bottleneck has shifted from the core to the
feeding path (Fig.~\ref{fig:capacity_fpga_control}).

Two challenges make this bottleneck fundamental.  First, the timing budget is
extremely tight: decomposition must finish within the $\mu$s-scale solve
window, even though clamping cost grows with problem size and graph density.
Second, prior work has focused on solver devices, problem formulations, and
decomposition
heuristics~\cite{9045100,IJNC148,8892209,11214007,Clasun2025,Johnson2011,Lo2023,Jiang2023},
with little guidance on sizing the digital hardware that feeds a
capacity-constrained solver.  Designers thus lack a simple way to answer three
questions: when a CPU controller will fail, how much hardware parallelism
keeps the solver utilized, and how much memory bandwidth a target problem
class needs.

The key insight of this work is that feeding cost is set by the $\rho N$
\emph{boundary} edges the clamp must fetch each iteration, a memory-bound
quantity that grows with problem size and is therefore out of reach of faster
arithmetic, multithreading, or vectorization.  The orchestration layer must be
\emph{sized}, not merely optimized.

We therefore propose an FPGA-ASIC co-design framework and a sizing methodology
for hybrid analog-digital Ising systems, built on three analytical quantities, $N^*$, $P_{\min}$, $Q_{\min}$, that translate core time, problem
density, and clock frequency into concrete hardware requirements.  Guided by
them we build an FPGA orchestration layer co-located with the chip, performing
low-latency breadth-first search (BFS) variable selection, clamping, and
subproblem construction, so the cores stay continuously fed with margin.
We state the scope explicitly: this work proposes no new Ising-solving
\emph{algorithm}: the clamped framework is qbsolv's~\cite{booth2017partitioning}
and the selection policy is BFS~\cite{Clasun2024,Clasun2025}.  To our knowledge
this is the first work to identify the orchestration layer as the dominant
bottleneck for capacity-constrained $\mu$s-class solvers, size it analytically,
and accelerate it on co-located silicon.

\subsection{Contributions}

\begin{itemize}
    \item \textbf{Chip design and fabrication.}
    A five-core coupled-oscillator Ising chip in 28~nm CMOS, each core
    supporting 45 all-to-all-coupled spins at 24~mW with core time
    $T_{\mathrm{core}} = 77.5~\mu$s.

    \item \textbf{Sizing methodology.}
    A methodology that treats the orchestration layer as a first-class design
    object: $N^*$ (critical scale) diagnoses when an orchestrator fails, while
    $P_{\min}$ and $Q_{\min}$ analytically fix the clamping and
    subproblem-generation parallelism needed to sustain full solver
    utilization for a given problem class, density, and clock frequency.

    \item \textbf{FPGA decomposer design.}
    An orchestration layer sized by the methodology and co-located with the
    chip, with dual-level parallelism: $P$ Processing Elements (PEs) for
    clamping and $Q$ Content-Addressable Memory (CAM) pairs for subproblem
    construction, plus a PCIe host interface and a 128-bit DDR3 AXI4 burst interface for problems beyond block RAM (BRAM) capacity.

    \item \textbf{Evaluation on fabricated silicon.}
    On-board measurement across coloring, MaxCut, and SAT shows
    $1.8\times$--$3.7\times$ pipeline margin, giving
    $9.8\times$--$46.5\times$ raw TTS improvement and $85\times$--$116\times$
    device-level power reduction over an optimized CPU orchestrator.
\end{itemize}

%% file: sections/2.background.tex
\section{Background}\label{sec:background}

\subsection{QUBO and the Ising Formulation}

Graph coloring, MaxCut, and SAT map to
Quadratic Unconstrained Binary Optimization (QUBO)~\cite{Lucas_2014},
minimizing $\mathbf{x}^{\top}Q\mathbf{x}$ over
$\mathbf{x}\in\{0,1\}^N$.  Ising machines operate on spins, and QUBO maps to
the Ising Hamiltonian via $s_i = 2x_i - 1$:
$H(\mathbf{s}) = -\sum_{i}\sum_{j} J_{ij} s_i s_j - \sum_{i} h_i s_i$, where $s_i \in \{-1,+1\}$ are spins, $h_i$ are local fields, and $J_{ij}$ are
couplings.  When the problem size $N$ exceeds the spin capacity $C$ (here
$C{=}45$), iterative decomposition is required~\cite{booth2017partitioning}.

\subsection{Coupled-Oscillator Ising Chip}

The chip is fabricated in 28~nm CMOS and implements an
all-to-all-coupled ring-oscillator Ising solver supporting 45 spins with
$\binom{45}{2} = 990$ programmable coupling edges ($-7$ to $+7$).
Each of the 990 couplers is an ultralow-power logic-based circuit with no
static power; measured total active power is 24~mW.  The core time
$T_{\mathrm{core}} \approx 77.5~\mu$s comprises program time, analog solving
time (multiple solves until termination), and readout.
Fig.~\ref{fig:cobi_layout} shows the die photograph.

\subsection{Decomposition Algorithm}\label{sec:formulation}

We adopt the decomposition framework in~\cite{booth2017partitioning}, which
partitions the problem into
overlapping hardware-sized subproblems submitted iteratively; spins outside the
selected subset are \emph{clamped} to their current values, folding their
influence into modified local fields.  The framework is agnostic to how the
subset is chosen: it uses an \emph{energy-impact} criterion, while
size-based~\cite{Okada2019} and multilevel~\cite{Ushijima-Mwesigwa2021,Angone2023}
methods trade preprocessing cost for solution quality.

This work keeps that clamped framework but replaces energy-impact selection
with \emph{BFS} selection~\cite{Clasun2024,Clasun2025}, which picks topologically connected
subsets of the QUBO graph on the principle that strongly coupled variables
should be optimized together.  On all-to-all-connected hardware such as the
target chip, this consistently outperforms energy-impact methods in quality
and convergence speed~\cite{Clasun2024,Clasun2025}.  Clamping and subproblem
construction are identical to qbsolv; only the selection policy differs.

\subsubsection{Problem Encoding}

Problems are encoded as Ising Hamiltonians before submission.  A 3SAT
instance with $n$ variables and $m$ clauses maps via the Chancellor
construction~\cite{Chancellor2016} (one ancilla per clause) to $N = n + m$
spins; \texttt{uf20} (20 variables, 91 clauses) gives $N = 111$.  Optimizing
SAT-to-Ising compilers~\cite{11617414} can reduce this encoding overhead; we
keep the standard construction so measured gains are attributable to the
decomposer alone.

\subsubsection{BFS Variable Selection}

BFS expands from a random seed until 45 spins are selected (one slot is
reserved for the local-field spin).  The seed is updated each iteration; on the
FPGA, randomization derives from the chip's returned spins.  BFS and Clamp are
sequential: BFS fixes the subgraph, then Clamp accumulates boundary-spin
influence (Eq.~(\ref{eq:clamp})).

\subsubsection{Clamped Subproblem Construction}

All spins outside $V_{\mathrm{sub}}$ are clamped to their current global
values.  The induced subproblem comprises $V_{\mathrm{sub}}$, the local
couplings $J_{ij}$ within it, and the modified local fields $h'_i$
(Eq.~(\ref{eq:clamp})).  It is transmitted to the chip, which returns
optimized spins $\mathbf{s}_{\mathrm{sub}}$; global states are updated for
$i \in V_{\mathrm{sub}}$ and the loop repeats until a satisfying assignment
is found or the iteration budget is exhausted.

%% file: sections/3.formulation.tex
\section{The Orchestration Bottleneck}\label{sec:analysis}

At \textmu s-scale core times the orchestration deadline becomes a hard
real-time constraint: every microsecond of feeding latency beyond core time
directly idles the solver.  This section derives the analytical structure of
the feeding cost and uses CPU profiling to show the gap is structural,
motivating the sizing methodology of Section~\ref{sec:architecture}.

\subsection{Decomposition Cost: Clamping and Subproblem Construction}

Each iteration selects a subset $V_{\mathrm{sub}}$ of $C{=}45$ spins and
splits the coupling-graph edges into \emph{internal} (both endpoints in
$V_{\mathrm{sub}}$) and \emph{boundary} (exactly one endpoint outside).
These two classes drive the two dominant decomposition costs.

\textbf{Clamping} accumulates the influence of external spins onto the
local fields of selected spins~\cite{booth2017partitioning}:
\begin{equation}
h'_i \;=\; h_i \;+\; \sum_{j \notin V_{\mathrm{sub}}} J_{ij}\, s_j^{(\mathrm{global})},
\quad \forall\, i \in V_{\mathrm{sub}}.
\label{eq:clamp}
\end{equation}
Each boundary edge requires one $J_{ij}$ lookup, and the total count is:
\begin{equation}
|E_{\mathrm{boundary}}| \;=\;
\sum_{i \in V_{\mathrm{sub}}} d^{\partial}_i
\;=\; |V_{\mathrm{sub}}| \cdot \bar{d}_{\mathrm{boundary}},
\label{eq:clamp_work}
\end{equation}
where $d^{\partial}_i = |\{j \notin V_{\mathrm{sub}} : J_{ij} \neq 0\}|$ is the
boundary degree of spin $i$.  At edge density $\rho$ each spin has degree
$\approx \rho N$, of which $\approx \rho C$ are internal ($C{=}45$), so:
\begin{equation}
\begin{split}
\bar{d}_{\mathrm{boundary}} &\;\approx\; \rho N - \rho C \;=\; \rho(N-C) \\
                             &\;\approx\; \rho N, \quad N \gg C,
\end{split}
\label{eq:dboundary}
\end{equation}
so $|E_{\mathrm{boundary}}| \propto \rho N$, increasing clamping cost
linearly with $N$.

\textbf{Subproblem construction (Subproblem)} assembles the $C \times C$ coupling
matrix by iterating over internal edges.  The internal-edge count is:
\begin{equation}
|E_{\mathrm{internal}}| \;\approx\; \rho_{\mathrm{sub}} \cdot \frac{C(C-1)}{2} \;\leq\; \binom{C}{2},
\label{eq:subq_work}
\end{equation}
where $\rho_{\mathrm{sub}}$ is the induced-subgraph density; since BFS selects
connected nodes, $\rho_{\mathrm{sub}} \geq \rho$ in general.  Subproblem cost
therefore depends on $C$ and $\rho_{\mathrm{sub}}$, not on $N$.  We write
$T_{\mathrm{clamp}}$ and $T_{\mathrm{Subproblem}}$ for the latencies of these
two stages and $T_{\mathrm{decomp}}$ for the total per-iteration feeding
latency (formalized in Eq.~(\ref{eq:tdecomp})).

\subsection{CPU Orchestration Profiling}

To show the feeding gap is structural, we profile the D-Wave Energy Impact
Decomposer (EID) pipeline~\cite{booth2017partitioning} on an Intel Core
i5-11500 with \texttt{perf} (1000 iterations), against a Compressed Sparse Row
(CSR) reimplementation that lays boundary edges contiguously
(Table~\ref{tab:cpu_llc}).

\begin{figure}[t]
\centering
\includegraphics[width=\linewidth]{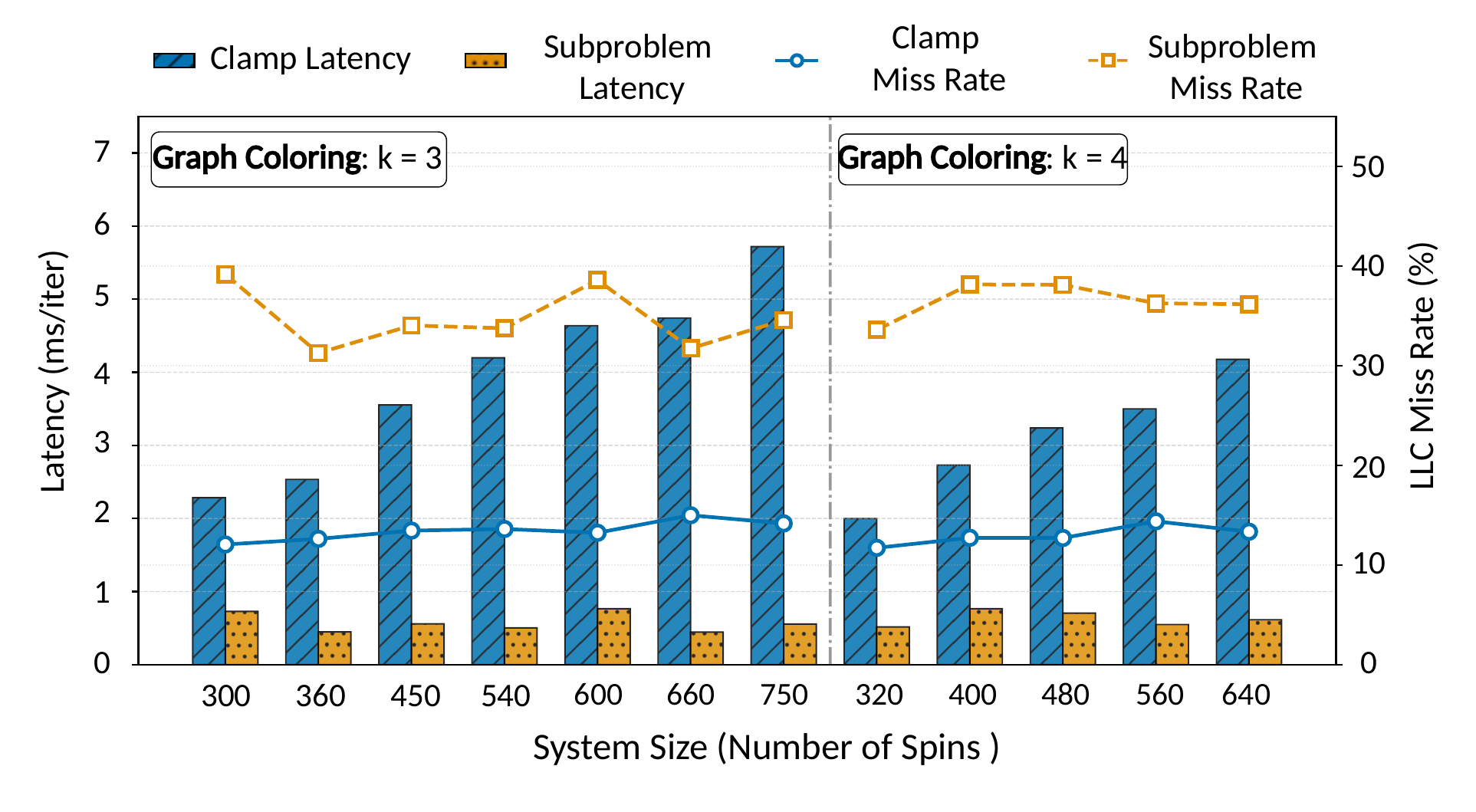}
\caption{Decomposition latency and LLC miss rate vs.\ system size (Coloring).
Clamp dominates and grows with $N$; Subproblem stays flat; miss rates stay
stable.}
\label{fig:coloring_latency_llc}
\end{figure}

The EID baseline sustains a Clamp Last-Level Cache (LLC) miss rate of
12--16\% across all three families, with no clear dependence on $N$.  CSR
reduces $T_{\mathrm{clamp}}$ on flat100 from 2.28~ms to 0.57~ms ($4.0\times$,
from sequential burst reads), yet the miss rate falls only from 12.1\% to
11.5\%: the boundary-edge working set spans the full graph and stays too large
for the LLC regardless of traversal order.
Clamp latency grows approximately linearly with $N$, consistent with
$|E_{\mathrm{boundary}}| \propto \rho N$.
Figure~\ref{fig:coloring_latency_llc} confirms this for coloring: Clamp latency
rises steeply with problem size while Subproblem latency stays roughly flat
(bounded by $C{=}45$) and the miss rate stays stable.

\begin{figure}[t]
\centering
\includegraphics[width=\linewidth]{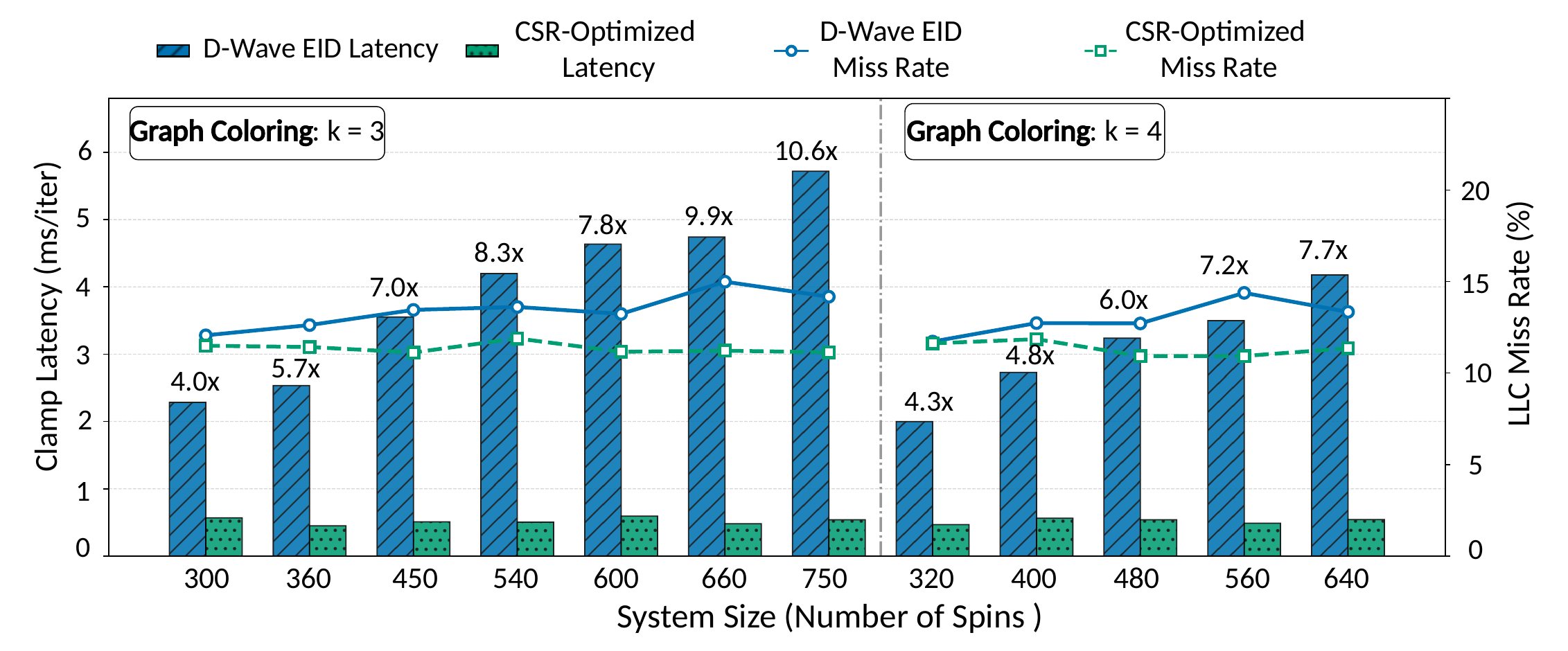}
\caption{Clamp latency vs.\ problem size: D-Wave EID vs.\ CSR-optimized
layout (Coloring).  CSR reduces latency by $4\times$, yet both remain far
above $T_{\mathrm{core}}=77.5~\mu$s.}
\label{fig:clamp_csr_vs_eid}
\end{figure}

Figure~\ref{fig:clamp_csr_vs_eid} shows that the CSR layout reduces Clamp
latency by $4\times$ over the EID baseline, yet both curves still far exceed
$T_{\mathrm{core}} = 77.5~\mu$s at every tested size.

\subsection{Quantifying the Orchestration Gap via $N^*$}

Table~\ref{tab:cpu_llc} summarizes LLC miss rates and latencies on flat100.
We define $N^* = N \cdot T_{\mathrm{core}}/T_{\mathrm{decomp}}$, the largest
problem size at which a given orchestrator still meets the solve window under
the measured linear scaling of feeding cost.

\begin{table}[t]
\centering
\caption{LLC profiling and $N^*$ on flat100 ($N{=}300$, k=3 coloring).
Both violate $T_{\mathrm{core}} = 77.5~\mu$s.}
\label{tab:cpu_llc}
\scriptsize
\setlength{\tabcolsep}{3pt}
\begin{tabular}{lcccccc}
\hline
Config & $T_{\mathrm{clamp}}$ & LLC\textsubscript{clamp} & $T_{\mathrm{Subproblem}}$ & LLC\textsubscript{Subproblem} & $T_{\mathrm{decomp}}$ & $N^*$ \\
\hline
D-Wave EID        & 2.28~ms & 12.1\% & 0.73~ms & 39.2\% & 3.01~ms & 8 \\
D-Wave EID CSR    & 0.57~ms & 11.5\% & 0.73~ms & 29.9\% & 1.30~ms & 18 \\
\hline
\end{tabular}
\end{table}

For the EID baseline, $T_{\mathrm{decomp}} = 3.01$~ms gives
$N^* = 300 \times 77.5/3010 = 7.7 \rightarrow 8$ nodes.  CSR reduces
$T_{\mathrm{clamp}}$ by $4.0\times$ but leaves $T_{\mathrm{Subproblem}}$
unchanged, so
$N^*_{\mathrm{CSR}} = 300 \times 77.5/1300 \approx 18$ nodes: $N^*$ improves
by only $2.2\times$ despite $4\times$ faster clamping.  Both configurations
violate the pipeline condition by $\geq 17\times$, motivating a purpose-built
orchestration layer.

%% file: sections/4.architecture.tex
\section{FPGA Architecture}\label{sec:architecture}

\begin{figure}[t]
\centering
\includegraphics[width=\linewidth]{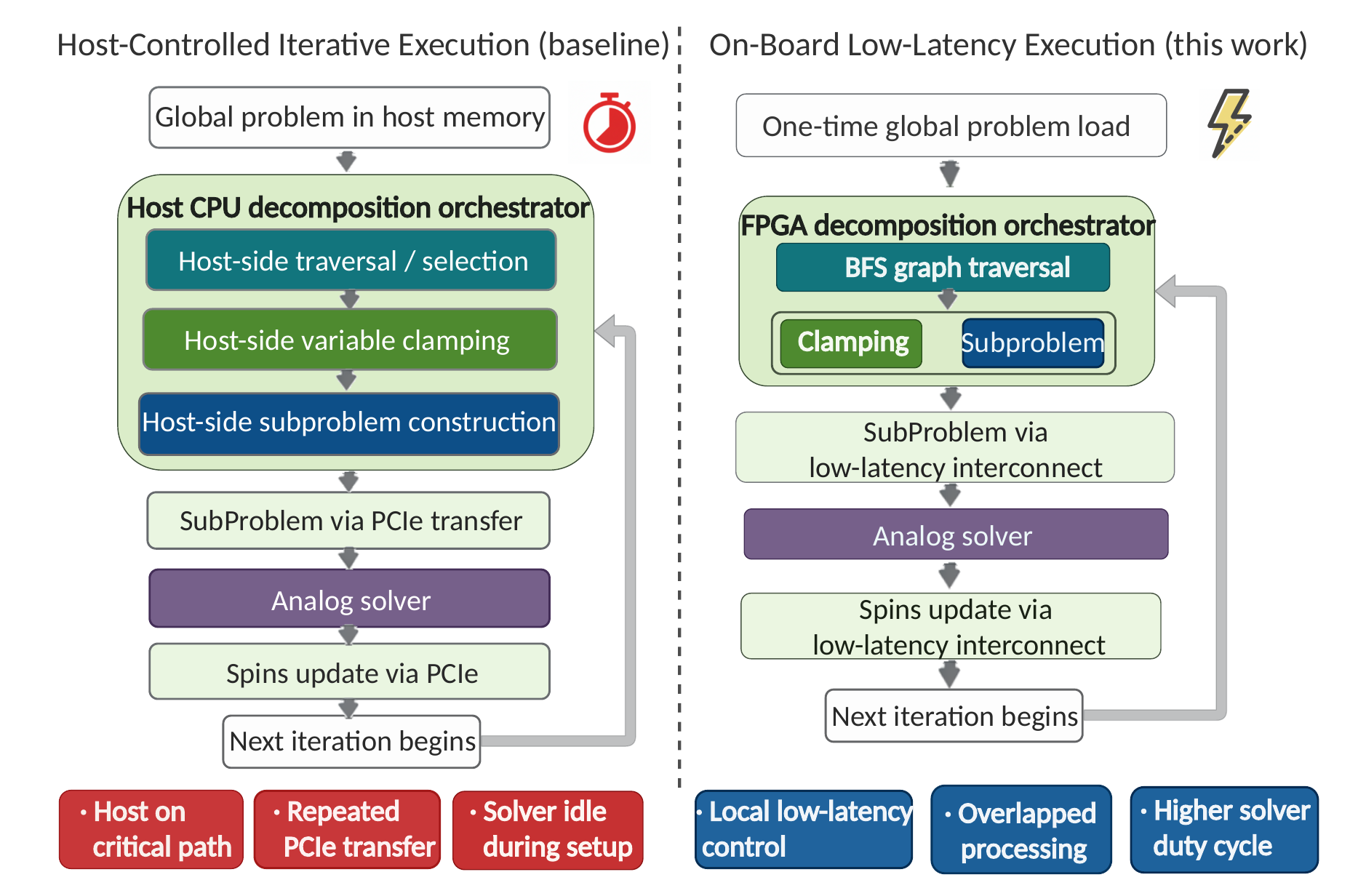}
\caption{Execution-flow comparison: moving decomposition from CPU to FPGA
removes PCIe transfers, increasing solver duty cycle.}
\label{fig:exec_flow_compare}
\end{figure}

We implement the orchestration layer on an FPGA co-located with the Ising
chip, sized by the methodology of Section~\ref{sec:analysis}.
Figure~\ref{fig:exec_flow_compare} contrasts the host-CPU baseline, where
repeated PCIe transfers add overhead and idle the solver, with the proposed
design, where decomposition runs entirely on the co-located FPGA.
Sections~\ref{sec:arch_clamp}--\ref{sec:arch_sizing} present, in turn, the
parallel clamping and subproblem-generation datapaths ($P$ PEs and $Q$ CAM
pairs), the five-stage pipeline and its overlap conditions, the board-level
energy advantage, and the design-knob equations for $P$, $Q$, and bandwidth.

\subsection{Hardware Design Advantage 1: Parallel Clamping and Subproblem Generation}\label{sec:arch_clamp}

\subsubsection{Memory Organization}

The global coupling matrix is stored in CSR
format~\cite{saad2003iterative} in on-chip BRAM.  CSR encodes only non-zero
entries, row pointers, column indices, and $J_{ij}$ laid out
contiguously, so all couplings of spin $i$ are fetched in one burst read.

\subsubsection{Parallel Processing Elements}

The $P$ PEs each process one CSR boundary-edge entry per clock cycle,
operating concurrently (Fig.~\ref{fig:clamping}(left)): neighbor-weight data for
each BFS-selected vertex is streamed from the shared AXI RDATA bus and
distributed across $P$ lanes, and an adder tree reduces the $P$ partial
products to one accumulated local field $h'_i$.  Since the PEs operate
simultaneously, the best-case clamping time is
\begin{equation*}
T_{\mathrm{clamp}} \geq |E_{\mathrm{boundary}}| / (f_{\mathrm{clk}} \times P),
\end{equation*}
where $f_{\mathrm{clk}} = 150$~MHz; the inequality reflects memory-access and
pipeline overhead beyond the ideal throughput limit.  This cost is
deterministic, free of the cache and OS-scheduling variability of the CPU
baseline.

\subsubsection{CSR Streaming for Subproblem Generation}

\begin{figure*}[t]
\centering
\begin{minipage}{0.45\linewidth}
\centering\includegraphics[width=\linewidth]{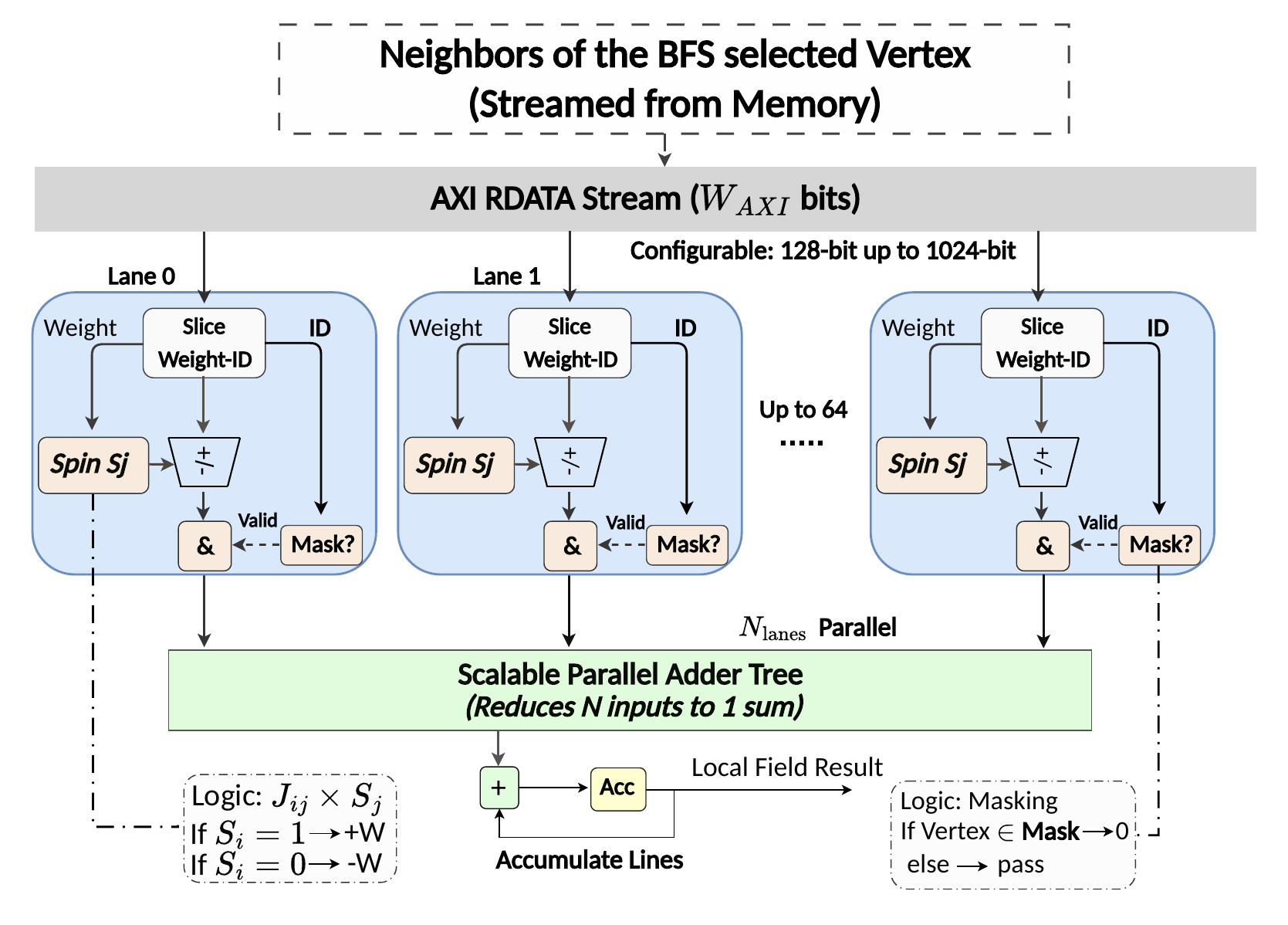}\\
\end{minipage}\hspace{0.03\linewidth}%
\begin{minipage}{0.45\linewidth}
\centering\includegraphics[width=\linewidth]{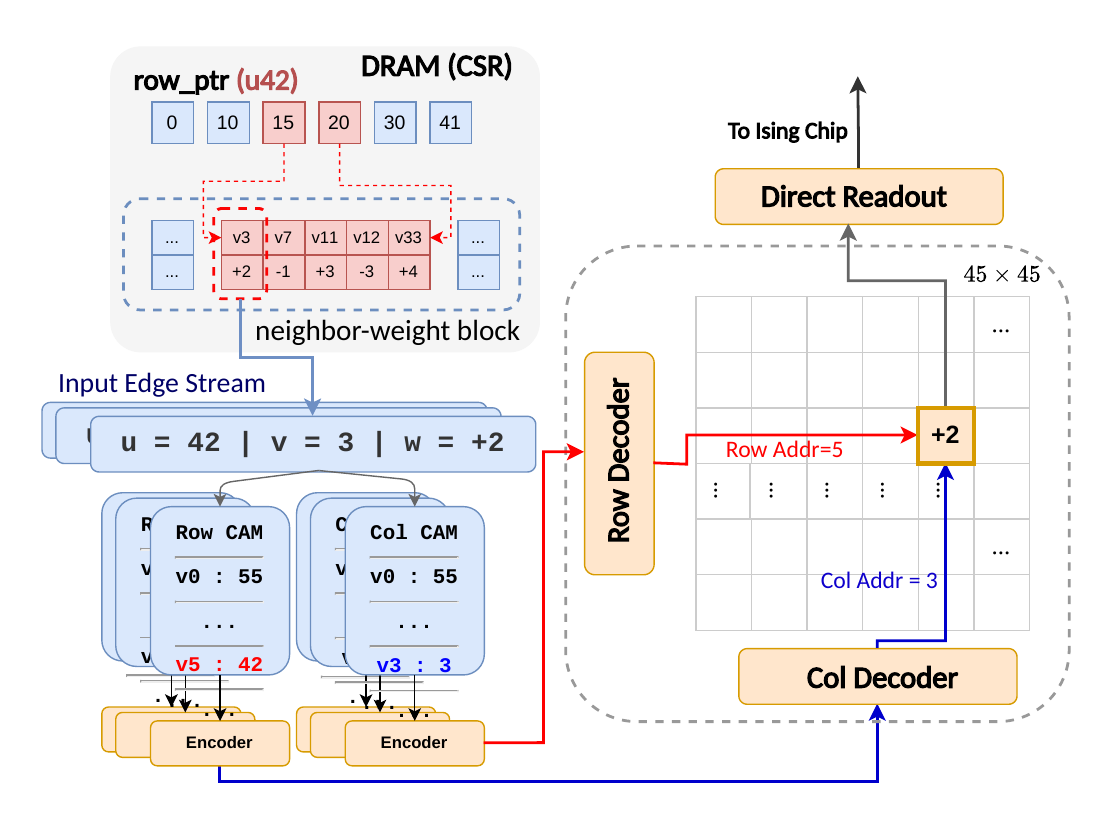}\\
\end{minipage}
\caption{(Left) Clamping unit: neighbor-weight data is streamed via AXI across
$P$ lanes, each computing $J_{ij} \cdot s_j$; an adder tree reduces the partial
products to one local-field accumulation, with masking for vertices already in
$V_{\mathrm{sub}}$.  (Right) Subproblem construction: each CSR edge $(u,v,w)$ is
resolved through Row/Column CAMs mapping global to local indices, writing $w$
into a $45 \times 45$ buffer read out to the chip.}
\label{fig:clamping}
\label{fig:subproblem_construction}
\end{figure*}

The Subproblem datapath extracts local couplings $J_{ij}$
for $(i,j) \in V_{\mathrm{sub}} \times V_{\mathrm{sub}}$ and assembles
the $45 \times 45$ coupling matrix transmitted to the Ising chip.
As Fig.~\ref{fig:clamping}(right) shows, each edge $(u, v, w)$ from the CSR
stream is dispatched to one of $Q$ parallel row/column CAM pairs, $Q$ being the
Subproblem parallelism degree (analogous to $P$).  DDR burst transfers
(128-bit AXI4) amortize access latency over coupling rows.

\subsection{Hardware Design Advantage 2: Pipeline Overlap with $T_{\mathrm{core}}$}\label{sec:arch_pipeline}

\subsubsection{Pipelined Decomposition Flow}

The five-stage pipeline is:

\textbf{Graph Traversal Unit (GTU)}: BFS over the CSR graph extracts a
  connected subgraph of up to $C = 45$ vertices, using AXI burst reads.
  \textbf{Clamp}: $P$ PEs compute the effective local fields $h'_i$
  (Eq.~(\ref{eq:clamp})) into a shared register file.
  \textbf{Subproblem Construction}: A parallel datapath extracts the
  induced subgraph, generating the local $J_{ij}$, concurrently with Clamp via
  decoupled FIFO streams.
  \textbf{Core}: $(J_{\mathrm{local}}, h'_{\mathrm{local}})$ is sent over
  the 16-bit shared bus; each chip returns the $\pm1$ assignment for the
  $C{=}45$ variables after $\approx$77.5~$\mu$s.
  \textbf{Feedback}: DFF register file updated,
  $s_i \leftarrow s_i^{(\mathrm{sub})}$ for $i \in V_{\mathrm{sub}}$.

\subsubsection{Pipeline Dependencies and Overlap}

Clamping and subproblem generation are independent RTL datapaths with no
shared mutable state, enabling two overlaps the sequential CPU orchestrator
could not exploit: BFS$(k{+}1)$ with Core$(k)$ + Feedback$(k)$ (no shared
data), and Clamp$(k)$ with Subproblem construction$(k)$ after BFS$(k)$
completes.  The per-iteration critical path is therefore:
\begin{equation}
\begin{split}
T_{\mathrm{iter}} &= \max(T_{\mathrm{GTU}},\, T_{\mathrm{core}} + T_{\mathrm{feedback}}) \\
                  &\quad + \max(T_{\mathrm{clamp}},\, T_{\mathrm{Subproblem}}),
\end{split}
\label{eq:titer}
\end{equation}

\textbf{What $T_{\mathrm{decomp}}$ measures.}
The feeding latency is the decomposition critical path
\begin{equation}
T_{\mathrm{decomp}} = T_{\mathrm{GTU}} + \max(T_{\mathrm{clamp}},\, T_{\mathrm{Subproblem}}).
\label{eq:tdecomp}
\end{equation}
It excludes
frame transmission and feedback, which belong to the solve path and overlap
$T_{\mathrm{core}}$, though the end-to-end TTS of Section~\ref{sec:eval_fpga}
does include them.  On board it is measured by a 32-bit counter at
$f_{\mathrm{clk}}$, started when decomposition is initiated and stopped at the
final
subproblem-output event.  Transmission is a fixed 663$\times$16-bit frame
(4.42~$\mu$s at 150~MHz), and only one frame is ever in flight, so the shared
bus adds no arbitration contention.

\subsection{Hardware Design Advantage 3: Energy Efficiency}\label{sec:arch_energy}

Energy improves for two reasons: the FPGA reduces orchestration runtime, and
board power stays low.  Over the Coloring ($k{=}3$) group
(Table~\ref{tab:aligned}) FPGA board power averages 0.204~W (0.173--0.245~W)
against a 22.8~W CPU group average (Running Average Power Limit, RAPL, net
above idle); evaluated at each
instance's own measured power, energy falls from 9.246~J to 3.13~mJ on flat100
($2950\times$) and 81.635~J to 31.6~mJ on flat200 ($2580\times$).  The CPU path stays
expensive because its core and memory hierarchy stay active throughout, which
the FPGA replaces with a deterministic dataflow.

\textbf{Scope of the power comparison.}
Both sides are reported at the \emph{device} level.  The
FPGA figure is board power for the orchestrator, PLL, and the 24~mW chip,
excluding PCIe and the host; in BRAM mode the DDR3 memory interface generator
(MIG) is not instantiated so its power is genuinely zero, and DDR adds its
measured
$\approx$0.71~W, keeping the FPGA path near 1~W.  The CPU figure is RAPL net
above idle, likewise excluding DRAM and board power, so a wall-plug comparison
would add the same host to both.

\subsection{Sizing the Interface: $P$, $Q$, and Bandwidth}\label{sec:arch_sizing}

The sizing methodology exposes three independent design parameters, each
directly derived from problem characteristics.

\textbf{Increasing $P$ (Clamp parallelism).}
The minimum PE count to satisfy the pipeline condition is:
\begin{equation}
P_{\min} = \left\lceil \frac{|E_{\mathrm{boundary}}|}{f_{\mathrm{clk}} \times T_{\mathrm{core}}} \right\rceil.
\label{eq:pmin}
\end{equation}
Since $T_{\mathrm{clamp}} \propto 1/P$, doubling $P$ halves clamping latency
at the cost of additional BRAM banks and PE logic.
Equation~(\ref{eq:pmin}) is a technology-independent \emph{throughput lower
bound}: it assumes every lane is used every cycle, whereas a real datapath
loses a fraction of each vertex's first and last fetched word to CSR row
alignment.  A practical implementation therefore provisions modest headroom
above the bound, $P_{\min,\mathrm{eff}} = P_{\min}/\eta$ with an
implementation efficiency $\eta \leq 1$; the deployed design carries
$1.8\times$--$3.7\times$ measured headroom (Table~\ref{tab:fpga_pipeline}),
which absorbs this loss.

\textbf{Increasing $Q$ (Subproblem parallelism).}
Each of the $Q$ CAM pairs handles one internal edge write per cycle, so
$T_{\mathrm{Subproblem}} \propto |E_{\mathrm{internal}}| / Q$, where
$|E_{\mathrm{internal}}|$ denotes the number of internal coupling edges within
$V_{\mathrm{sub}}$.  The minimum parallelism is
\begin{equation}
Q_{\min} = \left\lceil \frac{|E_{\mathrm{internal}}|}{f_{\mathrm{clk}} \times T_{\mathrm{core}}} \right\rceil.
\label{eq:qmin}
\end{equation}
The deployed design uses $Q{=}2$, the maximum that
fits within the available Artix-7 resources at $P{=}8$, providing
additional margin beyond the analytical minimum.

Higher $P$ and $Q$ require proportionally increased memory bandwidth to
sustain the data delivery rate.  Both are derived from $T_{\mathrm{core}}$,
$|E_{\mathrm{boundary}}|$, $|E_{\mathrm{internal}}|$, and $f_{\mathrm{clk}}$,
giving a first analytical estimate of the hardware before implementation.

%% file: sections/5.evaluation.tex
\section{Evaluation}\label{sec:evaluation}

Section~\ref{sec:eval_setup} describes the platform, benchmarks, and metrics.
Section~\ref{sec:eval_nstar} shows that the CPU feeding gap is structural and
survives every optimization we applied to the baseline.
Section~\ref{sec:eval_fpga} validates the sizing methodology on the physical
FPGA (pipeline condition, end-to-end TTS, solution quality, energy),
and Section~\ref{sec:eval_disc} discusses the broader implications.

\subsection{Experimental Setup}\label{sec:eval_setup}

\textbf{Hardware.}
FPGA: Xilinx Artix-7 XC7A35T (33,280 LUTs, 225~KB BRAM) at 150~MHz, co-located
with four Ising chips via a 16-bit shared input bus and individual 1-bit output
lines.  Each subproblem is sent as a fixed 663-word 16-bit frame (10,608
payload bits), and the traversal stage issues AXI burst reads of up to 16 beats
per row.  Benchmarks run in BRAM mode, with the DDR path (128-bit AXI4) as the
scalability option.
Post-implementation utilization for the largest instance (flat220, $N{=}660$,
$k{=}3$, $P{=}8$, $Q{=}2$) is 74.81\% LUT, 33.72\% DFF, 57.00\% BRAM.
CPU baseline: Intel Core i5-11500 @ 2.70~GHz via PCIe x4.

\begin{figure}[t]
\centering
\includegraphics[width=\linewidth]{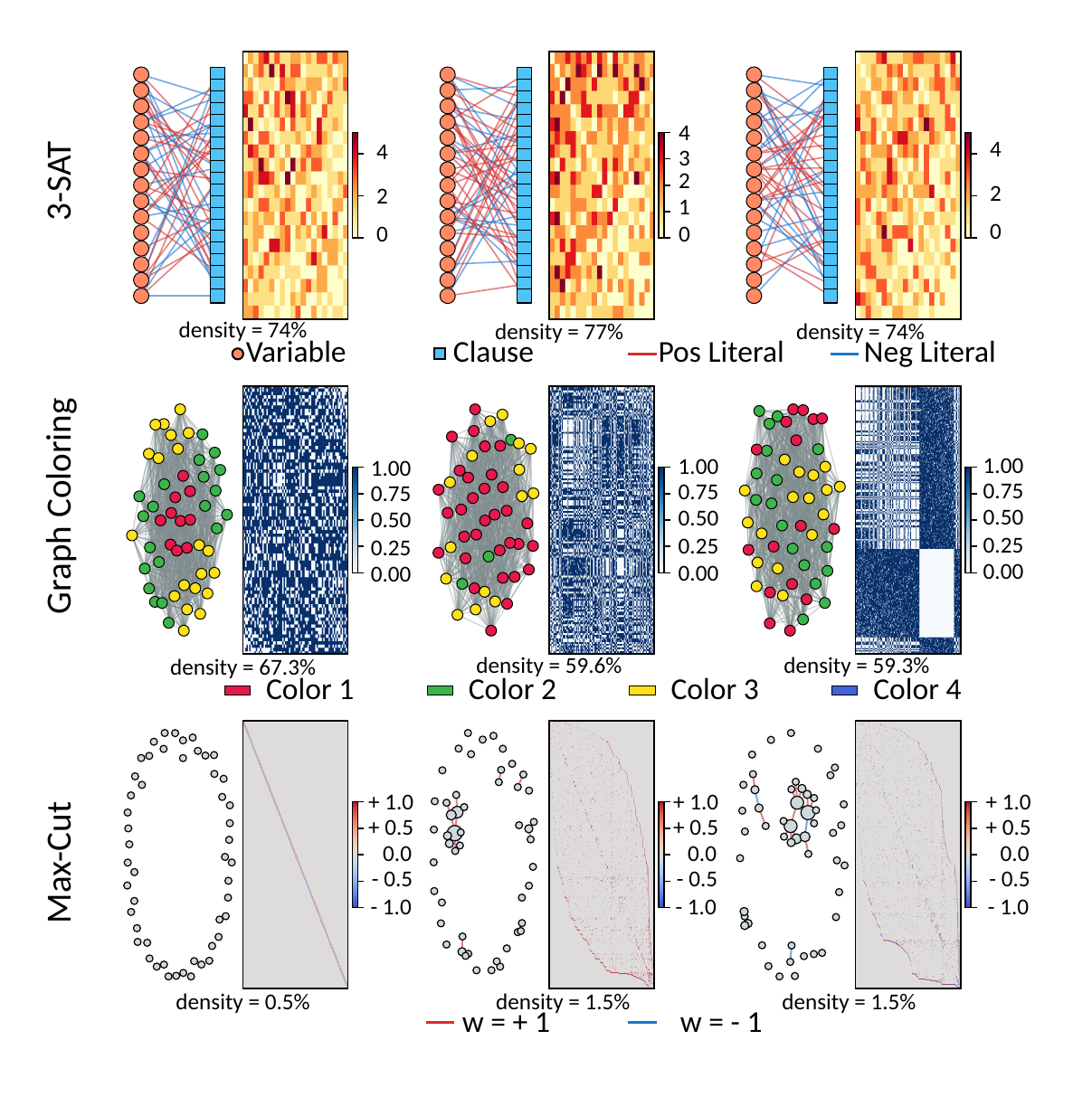}
\vspace{-10pt}
\caption{Coupling matrix structure of evaluated benchmarks.  3-SAT yields a dense
irregular matrix; Coloring produces block structure; MaxCut is sparse.}
\label{fig:benchmark_structure}
\end{figure}

\textbf{Benchmarks.}
Coloring instances (flat100--flat250) use a one-hot QUBO encoding, giving
$N{=}300$ to $N{=}750$ spins; MaxCut (G11--G21) maps directly with $N$ the
vertex count; 3SAT (uf20) uses the Chancellor
construction~\cite{Chancellor2016} from SATLIB~\cite{Hoos2000}.  All experiments use $\nu{=}100$ repeats of 500 iterations and
$p_{\mathrm{target}} = 0.99$.  Figure~\ref{fig:benchmark_structure} shows the
coupling structure of the three classes, 3-SAT dense and irregular, Coloring
block-diagonal, MaxCut sparse, differences that directly drive clamping cost
and subproblem density.

\textbf{Analog-core characterization.}
Since the system performs hundreds of repeated analog solves per instance,
Table~\ref{tab:chipchar} summarizes the measured core behavior the
orchestration layer must feed; per-node frequency-calibration bits and
sub-harmonic injection locking (SHIL) equalize the cores across corners.

\begin{table}[t]
\centering
\caption{Measured analog-core characterization (five 45-spin all-to-all cores
plus one reference, 28~nm, 0.9~V, 1.0$\times$1.7~mm).}
\label{tab:chipchar}
\small
\setlength{\tabcolsep}{3pt}
\begin{tabular}{p{1.9cm}p{5.9cm}}
\hline
Property & Measured value / mechanism \\
\hline
Coupling precision & 4-bit integer, $-7$ to $+7$ \\
Calibration & Per-node frequency-calibration bits + SHIL control bits, per core \\
Readout & 7-sample phase-sampling majority vote per spin \\
Vs.\ best-known & Against tabu best-known: 95\% energy target at
$\approx$100\% probability, 99\% at $\approx$80\%, exact ground state at
$\approx$60\% \\
Voltage / temp. & Stable across 0.7/0.9/1.1~V and $-40$/20/80~$^{\circ}$C \\
Power & 0.024~W per Ising core; digital post-processing $<$4\% of chip power and area \\
Solve stochasticity & Per-repeat iterations-to-success vary (uf20-01: mean 220.6, std 131.3, range 15--423) \\
\hline
\end{tabular}
\end{table}

\textbf{Metrics.}
We report $T_{\mathrm{decomp}}$, end-to-end time-to-solution (TTS), and
success rate (SR) on the chip, and energy.  TTS uses the standard
formula~\cite{booth2017partitioning}:
$\mathrm{TTS}=\frac{\ln(1-p_{\text{target}})}{\ln(1-p_{\mathrm{OI}})}
\times \bar{T}_{\mathrm{run}}$, with $p_{\mathrm{target}} = 0.99$ and
$p_{\mathrm{OI}}$ the per-run success probability, i.e.\ the
\emph{deployed} on-chip SR of Section~\ref{sec:eval_quality}.
\emph{Failure handling}: a repeat not reaching the target by the
500-iteration cap counts as a failure and contributes its full capped runtime,
lowering $p_{\mathrm{OI}}$ and raising $\bar{T}_{\mathrm{run}}$ at once, so raw
TTS already charges the FPGA twice for any SR deficit and we lead with it
throughout.  The secondary \emph{per-repeat-runtime ratio}
$\bar{T}_{\mathrm{CPU}}/\bar{T}_{\mathrm{FPGA}}$ removes the SR factor.
Success rates carry Wilson 95\% confidence intervals over $\nu{=}100$ repeats.
CPU energy uses Intel RAPL net-above-idle; FPGA energy = board power $\times$
TTS.

\subsection{Part I: $N^*$ Diagnostic Under CPU Orchestration}\label{sec:eval_nstar}

All results here use a C++ BFS decomposer, independent of (and not comparable
to) the Python EID pipeline of Section~\ref{sec:analysis}.

Table~\ref{tab:decomp_scaling} reports $T_{\mathrm{decomp}}$ and $N^*$ across
the Coloring~$k{=}3$ set.  On flat100 the BFS decomposer gives
$T_{\mathrm{decomp}} = 2.15$~ms ($27.7\times$ above $T_{\mathrm{core}}$,
$N^* = 10$); BFS+CSR reduces clamping $74\times$, raising $N^*$ to 117, but
0.20~ms still exceeds the deadline by $2.6\times$.  As $N$ grows BFS
$T_{\mathrm{decomp}}$ increases $11.2\times$, consistent with boundary-edge
count growing linearly in $N$; BFS+CSR is 10.8--23.3$\times$ faster but still
violates the condition at every size, and $N^*$ collapses from 117 to 56.

\subsubsection{Multithreaded and SIMD CPU Baselines}\label{sec:eval_cpuopt}

The baseline of Table~\ref{tab:decomp_scaling} is single-threaded and scalar.
Since clamping is embarrassingly parallel, we also evaluate OpenMP and AVX2
versions of it (Table~\ref{tab:cpuopt}) to test whether the gap is an
implementation artifact.

\begin{table}[t]
\centering
\caption{Optimized CPU baselines (i5-11500, Coloring $k{=}3$, 3 repeats
$\times$ 120 iters).  OpenMP: clamp speedup over 1 thread ($<$1 = regression).
AVX2: speedup over scalar, per stage, 1 thread.}
\label{tab:cpuopt}
\small
\setlength{\tabcolsep}{3pt}
\begin{tabular}{lcccc|cc}
\hline
\multirow{2}{*}{Instance ($N$)} & \multicolumn{4}{c|}{OpenMP clamp speedup} &
\multicolumn{2}{c}{AVX2 speedup} \\
 & 2T & 4T & 6T & 12T & Clamp & Subprob. \\
\hline
flat100 (300) & 1.38 & 0.83 & 0.75 & 0.67 & 1.69$\times$ & 1.00$\times$ \\
flat150 (450) & 1.14 & 0.75 & 0.78 & 0.70 & 3.06$\times$ & 1.00$\times$ \\
flat200 (600) & 1.19 & 0.89 & 0.82 & 0.87 & 3.44$\times$ & 0.96$\times$ \\
flat250 (750) & 1.23 & 1.07 & 1.07 & 1.03 & 3.62$\times$ & 0.94$\times$ \\
\hline
\end{tabular}
\end{table}

Neither brings $T_{\mathrm{decomp}}$ below $T_{\mathrm{core}}$: the cost is
structural, not computational.  Multithreading peaks at 1.14--1.38$\times$ at
two threads then \emph{regresses} to at best parity at 12, because the clamp is
memory-bound on the $\rho N$ boundary-edge working set, so extra threads
contend for the same cache hierarchy instead of adding throughput.  AVX2 vectorizes the clamp's
gather-multiply-accumulate for 1.69--3.62$\times$ but cannot reduce the
\emph{number} of $\rho N$ edges fetched; like the $4\times$ CSR gain it is a
one-time constant factor.  Subproblem construction is unmoved
($\approx$1.0$\times$): a data-dependent gather with no vectorizable
contiguous arithmetic, and BFS is inherently sequential.  Prefetching and GPU
offload do not help either: CSR already cuts $T_{\mathrm{clamp}}$
$4.0\times$ while moving the LLC miss rate only
12.1\%$\rightarrow$11.5\% (Table~\ref{tab:cpu_llc}), and a GPU is
latency-bound with one 45-spin subproblem per 77.5~$\mu$s window.  The gap is
a property of the feeding path, not of the baseline's engineering quality.

\setlength{\floatsep}{14pt plus 4pt minus 2pt}
\begin{table}[t]
\centering
\caption{C++ BFS decomposer $T_{\mathrm{decomp}}$ and $N^*$, Coloring~k=3
(i5-11500, $T_{\mathrm{core}} = 77.5~\mu$s).}
\label{tab:decomp_scaling}
\label{tab:ablation_cpu}
\small
\setlength{\tabcolsep}{3pt}
\begin{tabular}{lcrrrr}
\hline
Instance & $N$ & \multicolumn{2}{c}{BFS} & \multicolumn{2}{c}{BFS+CSR} \\
         &     & $T_{\mathrm{decomp}}$ & $N^*$ & $T_{\mathrm{decomp}}$ & $N^*$ \\
\hline
flat100  & 300 & 2.148~ms & 10  & 0.198~ms & 117 \\
flat120  & 360 & 3.780~ms & 7   & 0.266~ms & 104 \\
flat150  & 450 & 7.354~ms & 4   & 0.405~ms & 86  \\
flat180  & 540 & 11.015~ms & 3  & 0.538~ms & 77  \\
flat200  & 600 & 12.730~ms & 3  & 0.632~ms & 73  \\
flat220  & 660 & 15.950~ms & 3  & 0.773~ms & 66  \\
flat250  & 750 & 23.995~ms & 2  & 1.028~ms & 56  \\
\hline
\end{tabular}
\end{table}
\setlength{\floatsep}{4pt plus 1pt minus 1pt}

\subsection{Part II: Sizing Methodology Validation}\label{sec:eval_fpga}

\subsubsection{Pipeline Condition Satisfied}

Worst-case on-board $T_{\mathrm{decomp}}$ at $P{=}8$ is summarized below,
grouped by benchmark family.

\begin{table}[t]
\centering
\caption{Pipeline condition ($P{=}8$, $Q{=}2$, 150~MHz,
$T_{\mathrm{core}} = 77.5~\mu$s), worst case per benchmark group, measured
on-board on Artix-7 XC7A35T.  $^{\dagger}$DDR3-resident configuration
($P{=}8$, $Q{=}1$, 91\% LUT), board-level simulation with the real MIG; all
other rows are on-board measurements.}
\label{tab:fpga_pipeline}
\small
\setlength{\tabcolsep}{3pt}
\begin{tabular}{llc>{\centering\arraybackslash}p{2.35cm}}
\hline
Benchmark & Method & $T_{\mathrm{decomp}}$ ($\mu$s) &
\makecell{Condition\\(Pipeline Condition)} \\
\hline
uf20                      & BRAM & 21.1    & $\checkmark$ (3.7$\times$) \\
MaxCut G11--G21           & BRAM & 39.2    & $\checkmark$ (2.0$\times$) \\
Coloring flat100--flat180 & BRAM & 38.7    & $\checkmark$ (2.0$\times$) \\
Coloring flat200--flat250 & BRAM & 43.5    & $\checkmark$ (1.8$\times$) \\
uf20                      & DDR3$^{\dagger}$ & 74.8 & $\checkmark$ (1.04$\times$) \\
\hline
\end{tabular}
\end{table}

Table~\ref{tab:fpga_pipeline} shows the decomposer satisfies
$T_{\mathrm{decomp}} \leq T_{\mathrm{core}}$ everywhere, with headroom
$T_{\mathrm{core}}/T_{\mathrm{decomp}}$ of 1.8$\times$--3.7$\times$ even in
each group's worst case, eliminating solver idle periods.

\textbf{Beyond on-chip BRAM.}
BRAM mode is a deliberate resource-optimal choice: the MIG alone costs
$\approx$5.1K LUT plus PHY/IOs, so with the
orchestrator already at 74.81\% LUT it leaves no headroom for the $P$/$Q$
channels.  To demonstrate DDR-backed operation rather than assert it, we
place-and-route and timing-close a full DDR3 datapath (AXI FSM $\rightarrow$
MIG $\rightarrow$ MT41K256M16 DDR3-667, BL8) on the same Artix-7 at $P{=}8$,
$Q{=}1$ (91\% LUT).  Board-level simulation with the real MIG (150~MHz logic /
83.3~MHz DDR) shows a DDR-resident graph exceeding on-chip capacity decomposing
without stalling, the worst subproblem taking 74.8~$\mu$s
$< T_{\mathrm{core}}$; on board, the same configuration then runs
\texttt{uf20} entirely from DDR at an end-to-end TTS $\approx$21\% above BRAM
from off-chip latency, still far ahead of the CPU.  DDR is thus a validated
capacity extension costing a bounded latency premium.

\subsubsection{End-to-End TTS on the Physical Ising Chip}

Table~\ref{tab:e2e} reports end-to-end TTS on
the physical Ising chip for the FPGA orchestrator vs.\ the CPU BFS+CSR baseline
across all three benchmark classes.

\begin{table}[t]
\centering
\caption{End-to-end TTS (Coloring, SAT) and mean per-repeat runtime (MaxCut)
on the Ising chip.}
\label{tab:e2e}
\small
\setlength{\tabcolsep}{1pt}
\begin{tabular}{llcccccc}
\hline
Class & Instance & \multicolumn{2}{c}{CPU BFS+CSR} & \multicolumn{2}{c}{FPGA} & Raw & Per-rep. \\
      &          & TTS (ms) & SR & TTS (ms) & SR & Impr. & ratio \\
\hline
\multirow{7}{*}{\shortstack[l]{Coloring\\$k{=}3$}}
 & flat100 & 419.84  & 100\% & 18.11  & 90\% & 23.2$\times$ & 46.4$\times$ \\
 & flat120 & 1064.14 &  92\% & 24.55  & 81\% & 43.3$\times$ & 65.9$\times$ \\
 & flat150 & 2072.18 &  82\% & 131.81 & 72\% & 15.7$\times$ & 21.2$\times$ \\
 & flat180 & 4247.16 &  65\% & 192.18 & 65\% & 22.1$\times$ & 22.1$\times$ \\
 & flat200 & 3712.25 &  74\% & 129.16 & 68\% & 28.7$\times$ & 34.0$\times$ \\
 & flat220 & 5571.78 &  62\% & 142.72 & 65\% & 39.0$\times$ & 36.0$\times$ \\
 & flat250 & 7633.45 &  56\% & 271.72 & 56\% & 28.1$\times$ & 28.1$\times$ \\
\hline
\multirow{5}{*}{\shortstack[l]{Coloring\\$k{=}4$}}
 & flat80  & 1064.61 &  86\% &  45.99 & 83\% & 23.1$\times$ & 25.7$\times$ \\
 & flat100 &  814.87 &  95\% &  83.40 & 84\% &  9.8$\times$ & 16.0$\times$ \\
 & flat120 & 3025.30 &  74\% & 182.22 & 78\% & 16.6$\times$ & 14.8$\times$ \\
 & flat140 & 2933.63 &  76\% & 263.10 & 66\% & 11.2$\times$ & 14.8$\times$ \\
 & flat160 & 5430.50 &  64\% & 360.50 & 52\% & 15.1$\times$ & 21.0$\times$ \\
\hline
\multirow{10}{*}{SAT}
 & uf20-01  &  24.00  & 100\% &  27.70 &  87\% & 0.87$\times$ &  1.95$\times$ \\
 & uf20-02  &   9.42  & 100\% &   1.77 & 100\% & 5.32$\times$ &  5.32$\times$ \\
 & uf20-03  & 490.15  &  94\% &  52.81 &  91\% & 9.28$\times$ & 10.85$\times$ \\
 & uf20-04  &  79.22  & 100\% &  49.98 &  86\% & 1.59$\times$ &  3.72$\times$ \\
 & uf20-05  &  44.36  & 100\% &  34.16 &  94\% & 1.30$\times$ &  2.12$\times$ \\
 & uf20-06  &  76.41  & 100\% &  36.99 &  90\% & 2.07$\times$ &  4.13$\times$ \\
 & uf20-07  &  19.94  & 100\% &   0.77 & 100\% & 25.9$\times$ & 25.9$\times$  \\
 & uf20-08  &  80.18  & 100\% &  12.29 &  95\% & 6.52$\times$ & 10.03$\times$ \\
 & uf20-09  & 169.75  & 100\% &  39.54 &  91\% & 4.29$\times$ &  8.22$\times$ \\
 & uf20-010 &  19.19  & 100\% &  17.57 &  89\% & 1.09$\times$ &  2.29$\times$ \\
\hline
\multirow{2}{*}{MaxCut}
 & sparse (G11--G13) & 2364 & -- & 50.8 & -- & 46.5$\times$ & -- \\
 & dense  (G14--G21) & 1725 & -- & 48.4 & -- & 35.6$\times$ & -- \\
\hline
\end{tabular}
\end{table}

All FPGA TTS values include frame-transmission overhead.  For
Coloring~$k{=}3$, raw improvements span 15.7$\times$--43.3$\times$
(21.2$\times$--65.9$\times$ per-repeat), the gap largest when FPGA SR falls
below CPU SR; MaxCut delivers 35.6$\times$--46.5$\times$, sparse instances
(G11--G13) gaining more than dense (G14--G21) due to shorter per-subproblem
CPU processing.
For SAT, 8 of 10 instances show 1.09$\times$--9.28$\times$ raw
improvement.  Two instances (uf20-01, uf20-010) fall near 1$\times$ raw, and we qualify them
explicitly: the CPU already attains 100\% SR on these tiny instances (Wilson
95\% CI [96, 100]), leaving no headroom, while the FPGA reaches 87\% [79, 92]
and 89\% [81, 94].  Their per-repeat ratios (1.95$\times$, 2.29$\times$) show
the datapath advantage is intact and the margin grows with $N$.

\subsubsection{Success Rate and Solution-Quality Distributions}\label{sec:eval_quality}

The quality metric is benchmark-dependent: SR (fraction of repeats reaching a
valid coloring or satisfying all clauses) for Coloring and SAT, mean cut ratio
against the best-known reference for MaxCut.

\begin{figure}[t]
\centering
\includegraphics[width=\linewidth]{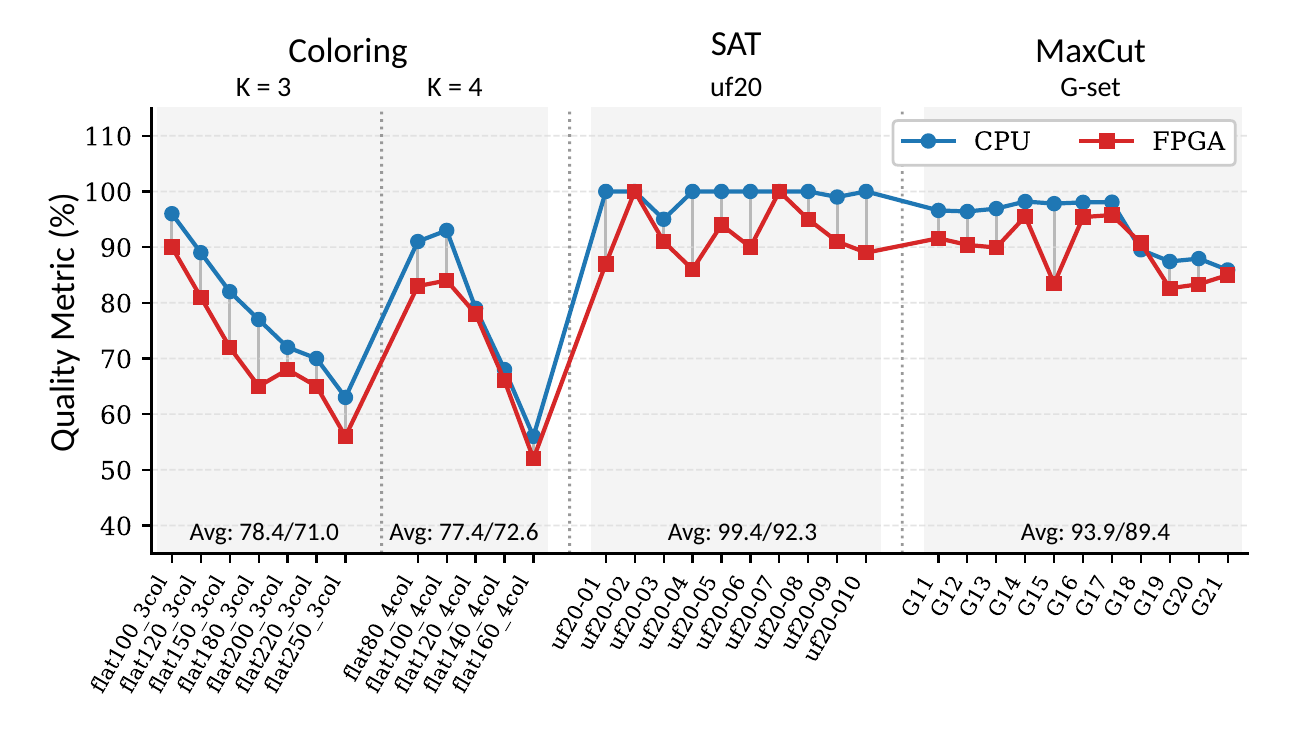}
\caption{Solution-quality distributions (SR for Coloring/SAT, cut ratio for MaxCut).}
\label{fig:solution_quality}
\end{figure}

The two paths do not induce identical stochastic trajectories
(Fig.~\ref{fig:solution_quality}): mean SR is 78.4\% (CPU) vs.\ 71.0\% (FPGA)
on Coloring~$k{=}3$ and 77.4\% vs.\ 72.6\% on $k{=}4$; for SAT both fully
satisfy when successful (best 91/91) but SR drops from 99.4\% to 92.3\%; for
MaxCut the mean cut ratio is 93.9\% vs.\ 89.4\%.

The gap originates in how BFS seeds are generated.  The CPU draws each seed
from a software random-number generator, whereas the FPGA reuses the chip's
returned spins, avoiding extra logic on a prototype already at 74.81\% LUT.  That stream is not fully independent: the NIST
SP800-22 spectral (DFT) test gives $p = 9.6\times10^{-7}$, rejecting the
i.i.d.\ null hypothesis, so the FPGA explores a narrower distribution of
decomposition trajectories.

\begin{table}[t]
\centering
\caption{(a)~Solution quality under synchronized BFS trajectories (10 repeats,
fixed seed; bold = FPGA $\geq$ CPU).  (b)~Board power group averages; CPU net
power = RAPL $\Delta$ above idle.}
\label{tab:aligned}
\small
\setlength{\tabcolsep}{3pt}
\begin{tabular}{lllcc}
\hline
\multicolumn{5}{l}{(a) Synchronized-trajectory quality} \\
\hline
Class & Instance & Metric & CPU & FPGA \\
\hline
\multirow{6}{*}{Coloring ($k{=}3$)}
 & flat120  & SR & 90\%  & \textbf{100\%} \\
 & flat150  & SR & 100\% & 100\% \\
 & flat180  & SR & 100\% & 100\% \\
 & flat200  & SR & 70\%  & \textbf{90\%} \\
 & flat220  & SR & 100\% & 100\% \\
 & flat250  & SR & 90\%  & \textbf{100\%} \\
\hline
\multirow{2}{*}{Coloring ($k{=}4$)}
 & flat80   & SR & 90\%  & \textbf{100\%} \\
 & flat100  & SR & 100\% & 100\% \\
\hline
\multirow{9}{*}{SAT}
 & uf20-02  & SR & 100\% & 100\% \\
 & uf20-03  & SR & 90\%  & 70\%  \\
 & uf20-04  & SR & 100\% & 90\%  \\
 & uf20-05  & SR & 100\% & 80\%  \\
 & uf20-06  & SR & 90\%  & 80\%  \\
 & uf20-07  & SR & 100\% & 100\% \\
 & uf20-08  & SR & 100\% & 100\% \\
 & uf20-09  & SR & 70\%  & \textbf{80\%}  \\
 & uf20-010 & SR & 100\% & 100\% \\
\hline
\multirow{2}{*}{MaxCut}
 & sparse (G11--G13) & Cut ratio & 99.3\% & 97.8\% \\
 & dense  (G14--G21) & Cut ratio & 96.4\% & \textbf{96.7\%} \\
\hline
\multicolumn{5}{l}{(b) Board power (group averages)} \\
\hline
Class & Instances & FPGA (W) & CPU (W) & Ratio \\
\hline
MaxCut           & G12--G21     & 0.230 & 21.5 & 94$\times$  \\
Coloring $k{=}3$ & flat100--220 & 0.204 & 22.8 & 111$\times$ \\
Coloring $k{=}4$ & flat80--160  & 0.190 & 22.6 & 119$\times$ \\
SAT              & uf20-01--010 & 0.174 & 22.4 & 128$\times$ \\
\hline
\end{tabular}
\end{table}

The synchronized-trajectory experiment (Table~\ref{tab:aligned}) isolates this
by forcing both paths to use the same seed at every iteration. This removes the Coloring gap entirely, with FPGA SR matching or exceeding the
CPU on every instance (flat200: 90\% vs.\ 70\%); SAT differs by at most 2/10
repeats and MaxCut ratios agree within 2\%.  The difference is therefore the
seed policy, not the decomposition datapath.  A hardware random-number
generator would fix it, but MT19937 needs 1--2 BRAMs and several hundred
LUTs~\cite{Konuma2005,Chandrasekaran2008}, competing with the $P$/$Q$ pipeline
on this Artix-7; a larger FPGA would favor adding one.  The penalty is in any
case already counted, since raw TTS uses the FPGA's measured SR.
FPGA board power (0.17--0.25~W) is 94--128$\times$ lower than CPU net
power (19.2--25.3~W).  Including the Ising chip's measured power
consumption of 24~mW, total power is 0.197--0.269~W (FPGA
path) versus 22.824~W (CPU path), a reduction of 85$\times$--116$\times$ at
the \emph{device} level, since both sides exclude the shared host
(Section~\ref{sec:arch_energy}).  Combined with the TTS reductions of
Table~\ref{tab:e2e}, taken at the same $p_{\mathrm{target}} = 0.99$, this
yields energy reductions of $2580\times$--$2950\times$ on representative
Coloring instances.

\textbf{Sensitivity to $P$ and $Q$.}
The sizing laws are validated by Table~\ref{tab:fpga_pipeline}: sizing from
problem characteristics selects $P{=}8$, $Q{=}2$, and that configuration meets
$T_{\mathrm{decomp}} \leq T_{\mathrm{core}}$ on real silicon for every
benchmark family.  Beyond that required operating point,
Table~\ref{tab:psweep} characterizes what further parallelism buys, sweeping
$P$ and $Q$ on a Kintex-7 XC7K160T (flat200, $N{=}600$, $k{=}3$,
post-implementation simulation; the Artix-7, already at 74.81\% LUT, cannot
host the larger points).

\begin{table}[t]
\centering
\caption{$P$/$Q$ sweeps (Kintex-7, post-implementation simulation).}
\label{tab:psweep}
\small
\setlength{\tabcolsep}{4pt}
\begin{tabular}{lcc|lcc}
\hline
$P$ & Clamp cyc. & Speedup & $Q$ & Subproblem cyc. & Speedup \\
\hline
8  & 2110 & 1.00$\times$ & 1 & 13214 & 1.00$\times$ \\
16 & 1440 & 1.46$\times$ & 2 &  8042 & 1.64$\times$ \\
32 & 1103 & 1.91$\times$ & 4 &  5474 & 2.41$\times$ \\
64 &  938 & 2.25$\times$ & 8 &  4200 & 3.15$\times$ \\
\hline
\end{tabular}
\end{table}

Latency keeps falling as either knob grows, with diminishing returns beyond
$P{=}16$ and $Q{=}4$: CSR row alignment wastes part of each vertex's first and
last fetched word, dispatch stalls on sparse vertices, and the fixed
subproblem read-out cannot be amortized by $Q$.  Provisioning past the deployed
point therefore buys margin rather than proportional speedup, which is why
Eqs.~(\ref{eq:pmin}) and~(\ref{eq:qmin}) are used as lower bounds with
headroom, not exact latency predictors.

\subsection{Discussion}\label{sec:eval_disc}

Across configurations $N^*$ tracks each design step: BFS gives $N^* = 10$; BFS+CSR raises it to 117, closing $11.7\times$ by data
layout alone but leaving $T_{\mathrm{decomp}}$ $2.6\times$ above
$T_{\mathrm{core}}$; FPGA CSR ($P{=}8$, $Q{=}2$) satisfies the condition at
every size.  The diagnostic thus identifies when the feeding path cannot keep
up, and $P_{\min}$/$Q_{\min}$ fix the required parallelism and memory bandwidth
before implementation begins.

\textbf{Core and chip utilization.}
All reported results use one active core per iteration, a property of the
algorithm: a BFS-qbsolv trajectory is strictly sequential, since subproblem
$k{+}1$ is clamped from the spins returned by subproblem $k$.  What the
orchestrator changes is that core's \emph{duty cycle}
$T_{\mathrm{core}}/T_{\mathrm{iter}}$: under CPU orchestration it stays below
2\% on flat250 ($\approx$7\% even charging PCIe as free), whereas the FPGA
hides decomposition behind the solve, lifting it to $\approx$62\% on flat250
and $\approx$75\% on \texttt{uf20}.
The remaining cores raise \emph{throughput}, not per-trajectory latency, by
running independent repeats concurrently: on the quad-chip 20-core card these
give 3.5--5.6$\times$, saturating at 7--8 instances at the shared host/PCIe
feeding path, further evidence that the feeding path is binding.

\textbf{TTS improvement attribution.}
Table~\ref{tab:attribution} attributes the flat250 ($k{=}3$, $N{=}750$)
per-iteration speedup to each architectural decision, one at a time. Kernel acceleration ($7.09\times$) dominates, exceeding co-location
($3.65\times$) and all pipeline overlap combined ($1.26\times$).
The realized TTS speedup
($28.1\times$) falls below the per-iteration figure ($32.6\times$) because the
FPGA needs more iterations per successful solve (290.9 vs.\ 204.9), a
$1.42\times$ trajectory-diversity penalty consistent with the fixed-seed
experiment (Table~\ref{tab:aligned}), not a hardware limitation.

\begin{table}[t]
\centering
\caption{Per-iteration speedup attribution (flat250, $k{=}3$, $N{=}750$).  Each
row adds one optimization; $T_{\mathrm{iter}}$ from average per-stage cycle
counts.}
\label{tab:attribution}
\small
\setlength{\tabcolsep}{3pt}
\begin{tabular}{lcrr}
\hline
Configuration & $T_{\mathrm{iter}}$ & Step & Cumul.\ \\
\hline
0.\ CPU, PCIe (baseline)           & 4.077~ms & --           & $1\times$ \\
1.\ CPU, PCIe overhead removed & 1.118~ms & $3.65\times$ & $3.65\times$ \\
2.\ FPGA, sequential (baseline)         & 0.158~ms & $7.09\times$ & $25.9\times$ \\
3.\ FPGA w/ Clamp$\|$Subproblem concurrency & 0.135~ms & $1.17\times$ & $30.3\times$ \\
4.\ FPGA w/ BFS-Core overlap (actual) & 0.125~ms & $1.08\times$ & $32.6\times$ \\
\hline
\end{tabular}
\end{table}

%% file: sections/6.conclusion.tex
\section{Conclusion}\label{sec:conclusion}

When a problem exceeds an analog Ising machine's spin capacity, a digital layer
needs to decompose, clamp, and feed subproblems within the solver's core time, and
at \textmu s-class core times it is that layer, not the analog core, that
limits throughput.  The gap is structural: it persists under every CPU
configuration we evaluated, including multithreaded and SIMD-optimized ones,
because the cost is data-dependent memory access rather than arithmetic.  The
methodology introduced here makes the constraint tractable, with $N^*$
diagnosing when an orchestrator fails and $P_{\min}$ and $Q_{\min}$ turning
core time, clock frequency, and problem structure into a required parallelism.
Sizing this way selects $P{=}8$, $Q{=}2$, and the resulting co-located Artix-7
orchestrator meets $T_{\mathrm{decomp}} \leq T_{\mathrm{core}}$ on real silicon
with $1.8\times$--$3.7\times$ margin across all three benchmark families,
delivering $9.8\times$--$46.5\times$ raw TTS improvement and
$85\times$--$116\times$ device-level power reduction.

What sets that cost is the edge structure of the decomposition: clamping scales
with the $\rho N$ boundary edges fetched each iteration, while subproblem
construction scales with the internal edges inside a hardware-sized subset.
The three evaluated families bracket the structural extremes of that space,
from dense irregular 3-SAT (74--77\%) through block-structured Coloring
(59--67\%) to sparse MaxCut (0.5--1.5\%), so the sizing laws are exercised
across genuinely different coupling structures.  Transferring them to another solver needs only its core time,
clock frequency, and the edge structure of its target workloads.  The
constraint moreover tightens as analog hardware improves: a chip with
$T_{\mathrm{core}} = 10~\mu$s demands an orchestration layer $8\times$ more
aggressive than the one validated here.  Capacity limits are the norm for
physics-based solvers, since oscillator, memristive, and qubit-limited
annealers all share them, though machines with sparse connectivity or very
different programming latencies will need their own constants.